\documentclass[a4paper]{jpconf}
\usepackage{graphicx}
\usepackage{bm}

\begin{document}

\title{Non-local electron transport and Coulomb effects in three-terminal metallic conductors}

\author{D.S. Golubev$^1$ and A.D. Zaikin$^{1,2}$}

\address{$^1$ Institut f\"ur Nanotechnologie,
Karlsruher Institut f\"ur Technologie (KIT), 76021 Karlsruhe, Germany\\
$^2$I.E. Tamm Department of
Theoretical Physics, P.N. Lebedev Physics Institute, 119991
Moscow, Russia}

\begin{abstract}
We demonstrate a close relation between Coulomb effects in non-local electron transport and non-local shot noise in three-terminal metallic conductors. Provided the whole structure is normal, cross-correlations in shot noise are negative and Coulomb interaction tends to suppress both local and non-local conductances of the system. The behavior of normal-superconducting-normal structures at subgap energies is entirely different.  In the tunneling limit non-local differential conductance of such systems are found to have an S-like shape and can turn negative at non-zero bias. At high transmissions crossed Andreev reflection yields positive noise cross-correlations and Coulomb anti-blockade of non-local electron transport.
\end{abstract}


\section{Introduction}

It is well known that discreteness of electron charge stays behind a number of fundamentally important physical phenomena such as, e.g., shot noise in mesoscopic
conductors \cite{BB} and Coulomb blockade of charge transfer in tunnel junctions \cite{SZ}. Exactly for this reason these two seemingly different phenomena turn out to be closely related to each other: Coulomb blockade is stronger in conductors
with bigger shot noise \cite{GZ01,LY1}. This fundamental relation was subsequently confirmed in experiments \cite{Pierre}. Later a close relation between shot noise and Coulomb blockade was also identified in hybrid normal-superconducting (NS) structures \cite{GZ09}, where doubling of elementary charge due to Andreev reflection becomes important at subgap energies.

Recently the same idea was extended \cite{GZ2010} to non-local effects in electron transport across three-terminal normal-superconducting-normal (NSN) systems where entanglement between electrons in different normal terminals can be realized. Non-local electron transport in such systems is determined by an interplay between elastic cotunneling (EC) and crossed Andreev reflection (CAR) and was recently investigated both experimentally \cite{Beckmann,Teun,Venkat,Basel,Beckmann2} and theoretically \cite{FFH,KZ06,GKZ}. While non-interacting theory predicts that CAR never dominates over direct electron transfer, both positive and negative non-local signals have been detected in a number of experiments \cite{Beckmann,Teun,Basel,Beckmann2}. Theoretically it was argued that CAR could prevail over EC in the presence of
Coulomb interactions \cite{LY} or an external ac field \cite{GlZ09}. Negative non-local conductance
was also predicted in interacting single-level quantum dots in-between normal and superconducting terminals \cite{Koenig}.

In Ref. \cite{GZ2010} we have already demonstrated that interaction effects in non-local transport and non-local shot noise in NSN systems are intimately related. This relation, however, turns out to be much more complicated than in the local case \cite{GZ01,LY1,GZ09}. The main reason for that is positive cross-correlations in shot noise which may occur in normal-superconducting hybrids \cite{BB,AD}. In NSN structures such positive noise cross-correlations were demonstrated theoretically \cite{Pistolesi,Melin,GZ2010} and experimentally \cite{Chandrasekhar}.
Note that this feature is specific to superconducting systems and is totally absent in normal ones  where cross-correlations of fluctuating currents are known to always be negative \cite{BB}. Hence, it would be interesting to extend our theory
of non-local electron transport in NSN systems in the presence of electron-electron interactions \cite{GZ2010} to normal conductors and compare the corresponding results derived for superconducting and normal structures.
This is the main goal of the present paper.

Our paper is organized as follows. In Sec. 2 we define our model and outline the key steps of the derivation of the effective action for our
three-terminal metallic structure. In Sec. 3 we re-formulate our results in terms of equivalent Langevin equations describing real time dynamics of fluctuating voltages and currents and demonstrate how the expressions for both local and non-local conductances in the presence of Coulomb interaction can be related to the corresponding shot noise correlators. Sec. 4 and 5 are devoted to the analysis of the effect of electron-electron interactions on the conductance matrix respectively
for superconducting and normal central electrodes. A brief summary of our main observations is presented in Sec. 6.

\begin{figure}
\begin{center}
\includegraphics[width=8cm]{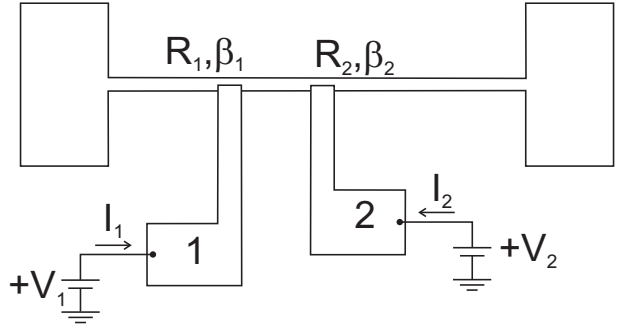}
\end{center}
\caption{Schematics of the system under consideration.}
\label{shema}
\end{figure}

\section{The model and effective action}

Let us consider a thin wire, which can be either normal or superconducting and is connected to two normal metallic leads via
two barriers. Quite generally, these barriers are characterized by two sets of channel transmissions $T_k^{(1)},T_k^{(2)}$.
For the sake of simplicity here we assume that the first junction has $N_1$ conducting channels with the same transmission
$T_1$, while the second one --- $N_2$ channels with transmission $T_2$.
Provided the wire is in the superconducting state, its energy spectrum has the gap $\Delta$. This gap, of course, equals to zero should the wire be in the normal state. In the latter case it is convenient to define the normal state conductances and the Fano factors of the barriers as follows
\begin{eqnarray}
G_{rr}^N = \frac{e^2}{\pi}N_rT_r,\;\;\beta_r^N =1-T_r,\;\; r=1,2.
\label{normal}
\end{eqnarray}
If the wire is superconducting, we define Andreev conductances and Andreev Fano factors,
which determine the transport properties across NS barriers in the subgap regime $eV_r,T\ll\Delta$:
\begin{eqnarray}
G_{rr}^{NS}=\frac{2e^2}{\pi}N_r\tau_r,\;\;\beta_r^{NS}=1-\tau_r,
\label{superconducting}
\end{eqnarray}
where $\tau_r=T_r^2/(2-T_r)^2$ represent effective Andreev transmissions of NS barriers.
Both barriers are supposed to have capacitances $C_1,C_2$ and
the leads 1 and 2 are characterized by large Ohmic conductances
$G_{1}^{\rm sh}$ and $G_{2}^{\rm sh}$. We also define the corresponding dimensionless
conductances of the electromagnetic environment
\begin{eqnarray}
g_r^{NS}=\frac{2\pi}{e^2}G_r^{\rm sh}.
\label{gNS}
\end{eqnarray}
The corresponding dimensionless conductances in case of a normal wire $g_r^N$ may differ
from $g_r^{NS}$ due to an additional contribution from the wire resistance.

Weak electromagnetic coupling between two
$NS$ barriers (e.g. via modes propagating in the superconductor \cite{LY}) will be disregarded.
Our main goal here is to evaluate electric currents $I_1(V_1,V_2)$ and $I_2(V_1,V_2)$ across the barriers 1 and 2.
Below we will consider two most interesting limits: (i) the subgap regime $T,eV_1,eV_2\ll\Delta$ in the case of a superconducting wire, and (ii) the regime of high energies or a normal wire.

The general Hamiltonian of our system can be expressed in the form \cite{GZ2010}
\begin{eqnarray}
H=H_1+H_2+H_{\rm wire}+H_{T,1}+H_{T,2},
\end{eqnarray}
where
$$
H_r=\sum_{\alpha=\uparrow,\downarrow}\int d{\bm x}
\hat \psi^\dagger_{r,\alpha}\left(-\frac{\nabla^2}{2m}-\mu\right)\hat \psi_{r,\alpha}, \;\;r=1,2,
$$
are the Hamiltonians of the normal metals, $m$ is electron mass, $\mu$ is the chemical potential,
$$
H_{\rm wire} = \int d{\bm x} \bigg[
\sum_{\alpha}\hat\chi^\dagger_{\alpha}\left(-\frac{\nabla^2}{2m}-\mu\right)\hat\chi_{\alpha}
+\Delta \hat\chi^\dagger_{\uparrow}\hat\chi^\dagger_{\downarrow}
+ \Delta^* \hat\chi_{\downarrow}\hat\chi_{\uparrow}
\bigg]
$$
is the Hamiltonian of a superconducting wire and
\begin{eqnarray}
H_{T,r}={\cal A}_r\sum_{\alpha,\beta=\uparrow,\downarrow}
\big[ t_r\, e^{i\varphi_r}\,  \hat\psi^\dagger_{\beta} \hat\chi_{r,\alpha}
+ t_r^*\, e^{-i\varphi_r}\,  \hat\chi_{r,\alpha}^\dagger\hat\psi_{\beta}  \big]
\label{HT}
\end{eqnarray}
are tunneling Hamiltonians describing transfer of electrons across the contacts with area ${\cal A}_r$ and tunneling amplitude $t_r$.
As we already indicated above, we will assume that both barriers are uniform
implying that all $N_r=k_F^2{\cal A}_r/4\pi$ conducting channels in the $r$-th barrier are characterized by equal transmission values
\begin{equation}
T_{r}={4\pi^2\nu_r\nu_{\rm wire}|t_{r}|^2}/{\left(1+\pi^2\nu_r\nu_{\rm wire}|t_{r}|^2\right)^2},
\label{ttr}
\end{equation}
where $\nu_j$ ($j=1,2$, wire) is the density of states in the corresponding terminal.
Finally, we note that fluctuating phases $\varphi_r$ introduced in Eq. (\ref{HT})
are linked to the voltage drops across the barriers $v_r$ by means of the standard relation $\dot\varphi_r = ev_r$ and are treated as quantum operators.

We will proceed in a standard manner and eliminate fermionic variables
expressing the kernel $J$ of the Keldysh evolution operator
via path integral over the phase fields \cite{SZ,deph}
\begin{equation}
J=\int \prod_{r=1,2}{\cal D} \varphi^F_r{\cal D}\varphi_r^B
\exp (iS_{\rm env}[\varphi^F_r,\varphi^R_r ]+iS_{T}[\varphi^F_r,\varphi^R_r ]), \label{pathint}
\end{equation}
where $\varphi^F_r$ and $\varphi_r^B$ are fluctuating phases defined respectively on the forward and backward branches of the Keldysh contour, $S_{\rm env}$ is the action of electromagnetic environment and the term $iS_{T}$ accounts for electron transfer between the terminals. In the case of linear Ohmic environment
considered here one has \cite{SZ}
\begin{eqnarray}
iS_{\rm env} = \sum_{r=1,2}\bigg[i\int dt \frac{(eV_{r}-\dot\varphi_r)(-C_r\dot\varphi_r^- +G_r^{\rm sh}\varphi^-_r)}{e^2}
-\frac{G_r^{\rm sh}}{2e^2}\int dt dt'\varphi^-_r(t)M(t-t')\varphi^-_r(t')\bigg],
\end{eqnarray}
where $\varphi_r=(\varphi_r^F+\varphi_r^B)/2$, $\varphi^-_r=\varphi_r^F-\varphi^B_r$
and
$$
M(t)=\int\frac{d\omega}{2\pi} e^{i\omega t}\omega\coth\frac{\omega}{2T}=-\frac{\pi T^2}{\sinh^2(\pi Tt)}.
$$
The term $iS_{T}$ is derived from the tunnel Hamiltonians (\ref{HT}) and reads
\begin{eqnarray}
iS_{T}={\rm tr}\ln  {\cal G}^{-1},\quad {\cal G}^{-1}=\left(
\begin{array}{ccc}
\check G_1^{-1} & \check t_1 & 0 \\
\check t_1^\dagger & \check G_{\rm wire}^{-1} & \check t_2 \\
0 & \check t_2^\dagger & \check G_2^{-1}
\end{array}\right),
\end{eqnarray}
where $4\times 4$ matrices $\check G_{j}^{-1}$ represent the inverse Keldysh Green functions
of isolated normal leads ($j=1,2$) and the wire, and $\check t_r$ is diagonal $4\times 4$ matrix
in the Nambu - Keldysh space
\begin{eqnarray}
\check t_r = \left(
\begin{array}{cccc}
-t_r e^{-i\varphi_r^F} & 0 & 0 & 0 \\
0 & t_r e^{-i\varphi_r^B} & 0 & 0 \\
0 & 0 & t_r e^{i\varphi_r^F} & 0 \\
0 & 0 & 0 & -t_r e^{i\varphi_r^B}
\end{array}
\right).
\label{t}
\end{eqnarray}
After some exact manipulations we obtain

\begin{eqnarray}
iS_{T} =
\,{\rm tr}\,\ln\big[ \check 1 - \check t_1^\dagger \check G_1\check t_1\check G_{\rm wire}
- \check t_2^\dagger \check G_2\check t_2 \check G_{\rm wire} \big].
\label{S}
\end{eqnarray}
While the expression (\ref{S}) for the action remains formally exact it is still too complicated to be directly employed in our calculations. In order to proceed we will make several additional steps which yield significant simplifications. First, we restrict ourselves to the limit of high conductances
\begin{equation}
g_r^{NS},\; g_r^N \gg 1,
\end{equation}
in which case phase fluctuations are weak and
it suffices to expand the action (\ref{S}) to the second order in $\varphi^-_r$, cf., e.g., \cite{GZ01,GZ09,Schmid}.
Next we assume that the resistance of the wire segment between the junctions, $r_{12}$, is small as compared
to the junction resistances, $r_{12}\ll 1/G_{rr}^N$. In this case one can additionally expand the action
in powers of the wire Green function connecting the two junctions.
In the case of a superconducting wire and at $T,eV_r\ll \Delta$ the corresponding calculation was elaborated in Ref. \cite{GZ2010}.
In this case the effective action of our system can be cast to the form
\begin{eqnarray}
iS_T=iS_{11}+iS_{22}+iS_{12},
\end{eqnarray}
where the contributions $iS_{11}$ and $iS_{12}$ read
\begin{eqnarray}
 iS_{11} = -i\frac{G_{11}}{e^2}\int dt \dot\varphi_1\varphi_1^-
-\int dtdt'\frac{\varphi^-_1(t)\,\tilde{\cal S}_{11}^{tt'}\,\varphi^-_1(t')}{2e^2},
\label{S11}
\end{eqnarray}
\begin{eqnarray}
iS_{12} = i\frac{G_{12}}{e^2}\int dt \,\big( \dot\varphi_1 \varphi_2^- + \dot\varphi_2 \varphi_1^- \big)
-\int dtdt'\frac{\varphi^-_1(t)\,\tilde{\cal S}_{12}^{tt'}\,\varphi^-_2(t')}{e^2}
\label{S12}
\end{eqnarray}
while the term $iS_{22}$ is obtained by interchanging the indices $1\leftrightarrow 2$ in Eq. (\ref{S11}). The functions $\tilde{\cal S}_{rl}^{tt'}$ are defined as
\begin{eqnarray}
\tilde{\cal S}_{11}^{tt'} &=&
G_{11}^{NS}M(t-t')\big( 1-\beta_1 +\beta_1\cos[2\varphi_1^{tt'}] \big)
+ 2G_{12}^{NS}M(t-t')\big( \alpha_1 - \eta_1\cos[2\varphi_1^{tt'}] \big)
\nonumber\\ &&
+(G_{12}^{NS}/2)M(t-t')\big( \kappa_1^+\cos[\varphi_1^{tt'}+\varphi_2^{tt'}]
+ \kappa_1^- \cos[\varphi_1^{tt'}-\varphi_2^{tt'}] \big),
\label{xi11}
\\
\tilde{\cal S}_{12}^{tt'} &=&
-G_{12}^{NS}M(t-t')\big( 1-\beta_1 +\beta_1\cos[2\varphi_1^{tt'}] \big)
-G_{12}^{NS}M(t-t')\big( 1-\beta_2 +\beta_2\cos[2\varphi_2^{tt'}] \big)
\nonumber\\ &&
+\,(G_{12}^{NS}/2)M(t-t')\big( \gamma_+\cos[\varphi_1^{tt'}+\varphi_2^{tt'}]
- \gamma_- \cos[\varphi_1^{tt'} - \varphi_2^{tt'}] \big).
\label{xi12}
\end{eqnarray}
Here we denoted $\varphi_r^{tt'}=\varphi_r(t)-\varphi_r(t')$.
Other parameters entering in Eqs. (\ref{xi11}) and (\ref{xi12}) read
\begin{eqnarray}
\alpha_r =\tau_r(1-2\tau_r)/\sqrt{\tau_1\tau_2},\hspace{2.4cm} &&
\eta_r =2\tau_r(1-\tau_r)/\sqrt{\tau_1\tau_2},
\nonumber\\
\kappa_r^{\pm} = \pm (4\tau_r-3) + 1/{\sqrt{\tau_1\tau_2}}\;\; (r=1,2), &&
\gamma_{\pm} = \pm 1 +\big({1-2\tau_1-2\tau_2+4\tau_1\tau_2}\big)/{\sqrt{\tau_1\tau_2}},
\label{kg}
\end{eqnarray}
while zero bias non-local conductance has the form
\begin{eqnarray}
G_{12}^{NS}=\frac{G_{11}^{NS}G_{22}^{NS}}{2e^2\nu_{\rm wire}} D(2i\Delta,\bm{r}_1,\bm{r}_2).
\end{eqnarray}
Here $D(\omega,\bm{r},\bm{r}')$ is the diffuson, which is defined as a solution of the following diffusion equation
\begin{eqnarray}
\left( -i\omega - D\nabla^2_{\bm{r}} \right)D(\omega,\bm{r},\bm{r}') = \delta(\bm{r}-\bm{r}').
\end{eqnarray}
In a simple
quasi-one-dimensional geometry of Fig. 1 one finds \cite{GKZ} $G_{12}^{NS}=G_{11}^{NS}G_{22}^{NS}R_\xi\,e^{-d/\xi}/2$, where $d$ is the distance between two $NS$ barriers
and $R_\xi$ is the resistance of the piece of the wire with the length equal to the superconducting coherence length $\xi=\sqrt{D/\Delta}$.
It is important to emphasize that {\it all} order terms in $t_r$ are fully accounted for in Eqs. (\ref{S11})-(\ref{kg}),
i.e. the action applies for arbitrary transmission values $T_{1,2}$ (or $\tau_{1,2}$) ranging from zero to one.

The calculation in the normal case goes along the same lines, however the resulting effective action turns out to be somewhat more
complicated because of retardation effects related to diffusion of electrons between the barriers. For the sake of simplicity here we avoid presenting an explicit form of the effective action in the normal case. Rather we will proceed directly to the final expressions for non-local currents which follow from this action. The corresponding expressions will be presented in Sec. 5.

\section{Langevin equations and interaction correction to the current}

It is well known that the quadratic in  $\varphi^-_{1,2}$ effective action can be exactly rewritten in terms of the corresponding Langevin equations \cite{Schmid,AES,GZ92} which describe the current balance in our system. In case of a superconducting wire these equations read \cite{GZ2010}
\begin{eqnarray}
C_1\dot v_1+(G_{1}^{\rm sh}+G_{11}^{NS})v_1-G_{12}^{NS}v_2=G_1^{\rm sh}V_1+\xi^{\rm sh}_1+\xi_1,
\nonumber\\
C_2\dot v_2+(G_{2}^{\rm sh}+G_{22}^{NS})v_2-G_{12}^{NS}v_1=G_2^{\rm sh}V_2+\xi^{\rm sh}_2+\xi_2.
\label{langevin}
\end{eqnarray}
Here $v_r$ are the fluctuating volatge drops across the junctions,
$\xi_r^{\rm sh}$ are stochastic variables with pair correlators
\begin{equation}
\langle\xi_r^{\rm sh}(t)\xi_r^{\rm sh}(t')\rangle=G_r^{\rm sh}M(t-t')
\end{equation}
describing Gaussian current noise in the shunt resistors,
while the variables $\xi_r$ with the correlators
\begin{equation}
\langle\xi_r(t)\xi_l(t')\rangle=\tilde{\cal S}_{rl}^{tt'}
\label{xiS}
\end{equation}
describe shot noise in NS barriers.

In case of normal wire the Langevin equation looks similar. One needs to replace $G_{rr}^{NS}$ by $G_{rr}^N$.
Besides that, the non-local conductance exhibits retardation effects, i.e.
one should replace, for example, $G_{12}^{NS}v_2 \to \int_{-\infty}^t dt' G_{12}^N(t-t')v_2(t')$.
Finally, the correlator of the noises also differs from (\ref{xiS}). As we already pointed out, the corresponding expressions are rather cumbersome and for this reason are not presented here.

Let us evaluate the current $I_1$ across the first barrier.
Solving Eqs. (\ref{langevin}) perturbatively in $1/g_r \ll 1$, in the lowest non-trivial order in this
parameter we get
\begin{equation}
I_1=G_{11}^{NS}V_1-G_{12}^{NS}V_2-\langle\xi_1\rangle.
\end{equation}
Here the average $\langle \xi_1 \rangle$ does not vanish since according to Eqs. (\ref{xi11}), (\ref{xi12}) the noise $\xi_1$
depends on the phases $\varphi_{1,2}$, which, in turn depend on $\xi_{1,2}$ by virtue of Eqs. (\ref{langevin}). Hence, we obtain
\begin{eqnarray}
\langle \xi_1 \rangle =
\left\langle \delta\varphi_1\;{\partial\xi_1}/{\partial\varphi_1}\right\rangle
+ \left\langle \delta\varphi_2\;{\partial\xi_1}/{\partial\varphi_2}\right\rangle,
\label{av1}
\end{eqnarray}
where the phase fluctuations $\delta\varphi_r$ are determined from Eqs. (\ref{langevin}). We obtain
\begin{eqnarray}
\delta\varphi_r(t)=e\int_{-\infty}^{t}dt' \frac{1-e^{-(t-t')/\tau_{RC}}}{G_r^{\rm sh}}\xi_r(t').
\end{eqnarray}
Here we assumed $G_{12}\ll G_{rr} \ll G_r^{\rm sh}$ and introduced the $RC-$time
$\tau_{RC}=C_r/G_{r}^{\rm sh}$. Substituting this expression into  Eq. (\ref{av1}) we get
\begin{eqnarray}
\frac{\langle \xi_1 \rangle}{e} =\sum_{r=1,2}\int_{-\infty}^{t}dt'
\frac{1-e^{-\frac{t-t'}{\tau_{RC}}}}{G_r^{\rm sh}}
\left.\frac{\partial\langle\xi_1(t)\xi_r(t')\rangle}{\partial\varphi_r(t)}\right|_{\varphi_r=eV_rt}.
\label{av}
\end{eqnarray}

Below we will make use of the above expressions and directly
evaluate the non-local currents in both interesting limits of superconducting and normal wires.

\section{Superconducting wire}

\begin{figure}[!h]
\begin{center}
\includegraphics[width=12cm]{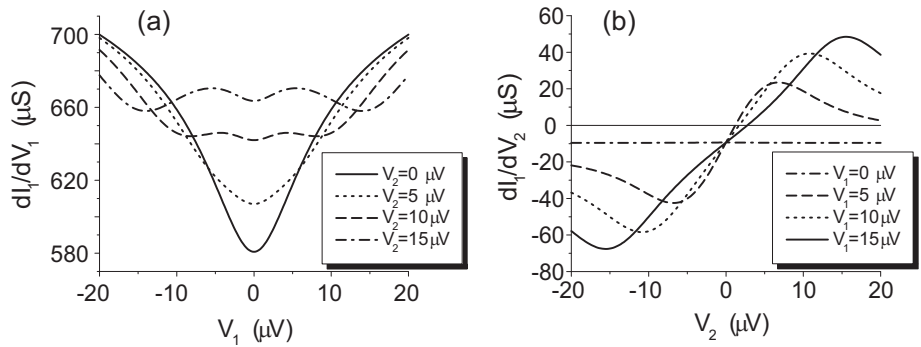}
\end{center}
\caption{Local (a) and non-local (b) differential conductances of a sample with superconducting wire, which are
defined, respectively, in Eqs. (\ref{di1dv1}) and (\ref{di1dv2}).
The parameters of the system are: $T=20$ mK, $G_{11}^{NS}=1$ mS, $G_{22}^{NS}=3.2$ mS, $G_{12}^{NS}=10$ $\mu$S, $g_1^{NS}=g_2^{NS}=516$,
$\tau_{RC}=10^{-12}$ s.
The  transmissions of the junctions are small, $T_1=0.063$, $T_2=0.11$, the corresponding numbers of channels are,
on the contrary, large, $N_1=N_2=6.1\times 10^6$. }
\label{fig_Gsuper}
\end{figure}

We begin with a superconducting wire. In this case employing
Eqs. (\ref{xi11}), (\ref{xi12}) and performing the time integral in Eq. (\ref{av}) we arrive at the following expression for
the current through the first barrier
\begin{eqnarray}
 I_1 &=& G_{11}V_1-G_{12}V_2
-\frac{2 G_{11}\beta_1-4G_{12}\eta_1}{g_1^{NS}}F_0(2V_1)
+ \frac{2G_{12}\beta_2}{g_2^{NS}}F_0(2V_2)
\nonumber\\ &&
-\, G_{12}\left( \frac{\kappa_1^+}{g_1^{NS}} + \frac{\gamma_+}{g_2^{NS}} \right)F_0(V_1+V_2)
- G_{12}\left( \frac{\kappa_1^-}{g_1^{NS}} + \frac{\gamma_-}{g_2^{NS}} \right) F_0(V_1-V_2),
\label{i1}
\end{eqnarray}
where
\begin{eqnarray}
F_0(V)=\,{\rm Re}\,\left[
- V\Psi\left(1+i\frac{eV}{2\pi T}\right)
+\left( V-\frac{i}{e\tau_{RC}} \right)\Psi\left(1+\frac{1}{2\pi T\tau_{RC}}+i\frac{eV}{2\pi T}\right)
\right].
\label{F0}
\end{eqnarray}
Taking the derivatives of the current  (\ref{i1}) with respect to $V_1$ and $V_2$ we define respectively local and non-local differential conductances of the first barrier. They read
\begin{eqnarray}
\frac{\partial I_1}{\partial V_1} &=& G_{11}-\frac{4 G_{11}\beta_1-8G_{12}\eta_1}{g_1^{NS}}F(2V_1)
\nonumber\\ &&
-\,G_{12}\left( \frac{\kappa_1^+}{g_1^{NS}} + \frac{\gamma_+}{g_2^{NS}} \right)F(V_1+V_2) -G_{12}\left( \frac{\kappa_1^-}{g_1^{NS}} + \frac{\gamma_-}{g_2^{NS}} \right)F(V_1-V_2),
\label{di1dv1}
\end{eqnarray}
and
\begin{eqnarray}
\frac{\partial I_1}{\partial V_2} &=& -G_{12}\left[1-\frac{4\beta_2}{g_2^{NS}}F(2V_2)\right]
\nonumber\\ &&
-\,G_{12}\left( \frac{\kappa_1^+}{g_1^{NS}} + \frac{\gamma_+}{g_2^{NS}} \right)F(V_1+V_2) + G_{12}\left( \frac{\kappa_1^-}{g_1^{NS}} + \frac{\gamma_-}{g_2^{NS}} \right)F(V_1-V_2),
\label{di1dv2}
\end{eqnarray}
where
\begin{eqnarray}
F(V) &=&\,{\rm Re}\,\left[ \Psi\left(1+\frac{1}{2\pi T\tau_{RC}}+i\frac{eV}{2\pi T}\right)
+ \left(\frac{1}{2\pi T\tau_{RC}}+i\frac{eV}{2\pi T}\right)\Psi'\left(1+\frac{1}{2\pi T\tau_{RC}}+i\frac{eV}{2\pi T}\right)
\right.
\nonumber\\ &&
\left.
-\, \Psi\left(1+i\frac{eV}{2\pi T}\right)- i\frac{eV}{2\pi T}\Psi'\left(1+i\frac{eV}{2\pi T}\right)\right],
\label{psi}
\end{eqnarray}
and $\Psi(x)$ is the digamma function. The conductances (\ref{di1dv1}), (\ref{di1dv2}) are displayed in Fig. \ref{fig_Gsuper}.

We observe that both these conductances are affected by Coulomb interaction which yields non-trivial corrections to the corresponding non-interacting expressions. In the interaction correction to the local conductance in Eq. (\ref{di1dv1}) we recover the Coulomb blockade term \cite{GZ09} $\propto \beta_1$ and, in addition, three non-local contributions. The first of them $\propto \eta_1$ enhances the conductance, while the second one
$\propto \kappa_1^+,\gamma_+$ provides additional Coulomb suppression of $\partial I_1/\partial V_1$.
The last term $ \propto \kappa_1^-,\gamma_-$ can be both positive (at $\tau_{1,2}\ll 1$) and negative (at bigger $\tau_{1,2}$) implying the tendency
to Coulomb {\it anti-blockade} in the latter case. The first term $\propto \beta_2$ in Eq. (\ref{di1dv2}) for the non-local conductance has an opposite sign as compared
to $G_{12}$ (thus implying Coulomb blockade), while the second one $\propto \kappa_1^+,\gamma_+$ yields Coulomb {\it anti-blockade}.
Finally, the third term $ \propto \kappa_1^-,\gamma_-$ tends to suppress or enhance the absolute value of the non-local conductance respectively
for ${\kappa_1^-}/{g_1^{NS}} + {\gamma_-}/{g_2^{NS}}>0$ and ${\kappa_1^-}/{g_1^{NS}} + {\gamma_-}/{g_2^{NS}}<0$.

As we already argued \cite{GZ2010} these non-trivial features of different Coulomb corrections are directly related to the corresponding contributions to shot noise. In particular, negatively and positively correlated noise terms are associated respectively with Coulomb blockade and anti-blockade terms in the above expressions for both local and non-local conductances.

\section{Normal wire}

Let us now turn to the case of a normal metallic wire. In this case the current $I_1$ takes the form
\begin{eqnarray}
I_1 = G_{11}^N V_1 - G_{12}^{N} V_2 - \frac{2 G_{11}^N\beta_1^{N} }{g_1^N} F_0(V_1)
+  \frac{2 G_{12}^{N}\beta_2^{N}}{g_2^N}   F_0(V_2)
-  \frac{2 G_{12}^{N} \beta_1^{N}}{g_1^N} F_0(V_1-V_2),
\end{eqnarray}
where the normal state non-local conductance is defined as follows
\begin{eqnarray}
G_{12}^{N} = \frac{D(0,\bm{r}_1,\bm{r}_2)}{2e^2\nu_{\rm wire}}G_{11}^NG_{22}^N.
\end{eqnarray}
Accordingly the differential resistances read
\begin{eqnarray}
\frac{\partial I_1}{\partial V_1} &=& G_{11}^N - \frac{2 G_{11}^N\beta_1^{N} }{g_1^N} F(V_1)
-  \frac{2 G_{12}^{N} \beta_1^{N}}{g_1^N} F(V_1-V_2),
\label{di1dv1_norm}
\end{eqnarray}
\begin{eqnarray}
\frac{\partial I_1}{\partial V_2} &=& -G_{12}^{N} + \frac{2 G_{12}^{N}\beta_2^{N} }{g_2^N} F(V_2) +  \frac{2 G_{12}^{N} \beta_1^{N}}{g_1^N} F(V_1-V_2).
\label{di1dv2_norm}
\end{eqnarray}
Here the functions $F_0(V)$ and $F(V)$ are again defined in Eqs. (\ref{F0}) and (\ref{psi}).

\begin{figure}[!h]
\begin{center}
\includegraphics[width=15cm]{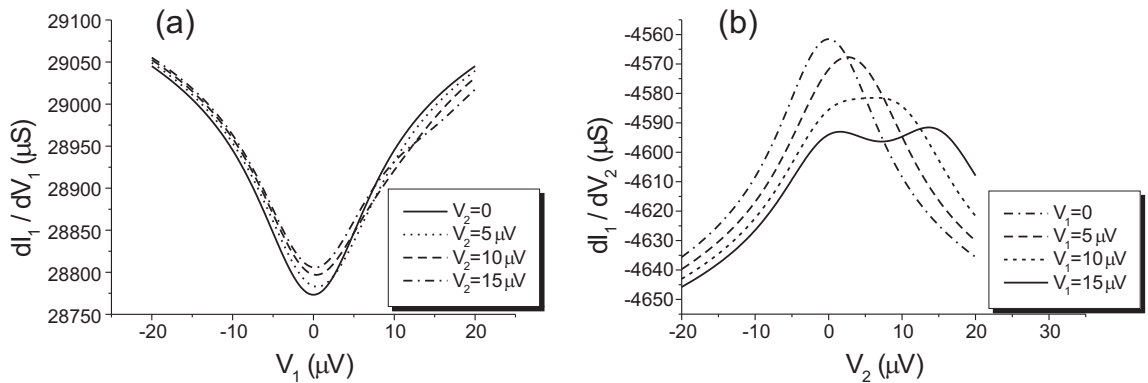}
\end{center}
\caption{Local (a) and non-local (b) differential conductances of a sample with normal wire, which are
defined, respectively, in Eqs. (\ref{di1dv1_norm}) and (\ref{di1dv2_norm}).
The system parameters are the same as in Fig. \ref{fig_Gsuper}, namely
$T=20$ mK, $g_1^N=g_2^N=516$, $T_1=0.063$, $T_2=0.11$, $N_1=N_2=6.1\times 10^6$, $\tau_{RC}=10^{-12}$ s.
Accordingly, the normal conductances take the values
$G_{11}^{NS}=29.8$ mS, $G_{22}^{NS}=52$ mS, $G_{12}^{NS}=4.83$ $\mu$S.  }
\label{fig_Gnormal}
\end{figure}

The conductances (\ref{di1dv1_norm}) and (\ref{di1dv2_norm}) are depicted in Fig. \ref{fig_Gnormal}.

Comparing these results with those obtained above for NSN systems we observe striking differences between them. The main difference is due to
the fact that correlations in non-local shot noise in normal multi-terminal systems are always negative, while positive cross-correlations may occur in structures involving superconductors. Accordingly, in the normal case Coulomb interaction tends to always suppress non-local conductance, and the corresponding interaction correction in Eq. (\ref{di1dv2_norm}) depends on the voltage difference $V_1-V_2$. In contrast, the expression for the non-local conductance in three-terminal NSN systems (\ref{di1dv2}) contains Coulomb terms which
depend both on $V_1-V_2$ and $V_1+V_2$ originating respectively from
negative and positive cross-correlations in shot noise. As we already discussed, these terms describe respectively Coulomb blockade and Coulomb
anti-blockade of the non-local conductance in the superconducting case.

\section{Discussion and summary}

In this paper we developed a theory elucidating a non-trivial physical relation between shot noise and Coulomb effects in non-local electron transport in three-terminal metallic structures. We evaluated non-local current-current correlators in such systems at arbitrary interface transmissions and arbitrary frequencies and directly related them to Coulomb effects in non-local electron transport.

One of the important features of NSN systems under consideration is that in the tunneling limit almost no
effect of Coulomb interactions on non-local conductance is expected if one of the applied voltages, $V_1$ or $V_2$, equals to zero. This effect is directly related to the cancellation between EC and CAR contributions to shot noise in the corresponding limit \cite{Pistolesi}. For nonzero $V_1$ and $V_2$  no such cancellation exists anymore and the non-local conductance $\partial I_1/\partial V_2$ approaches the S-like shape being enhanced at $V_1\approx V_2$ and partially suppressed at $V_1\approx -V_2$, see Fig. \ref{fig_Gsuper}b. Both these features have a clear physical interpretation. Indeed, at $V_1\approx -V_2$ negative cross-correlations due to EC dominate non-local shot noise leading to Coulomb blockade of non-local conductance while at $V_1\approx V_2$ positive cross-correlations due to CAR prevail and Coulomb anti-blockade of non-local transport is observed. At higher interface transmissions only Coulomb anti-blockade of non-local conductance remains, which is again related to CAR-induced positive cross-correlations in shot noise.

Comparing these results with those obtained in systems with normal central electrodes  we observe
striking differences. In particular, both local and non-local conductances always tend to be suppressed by Coulomb interaction and anti-blockade
effects never occur in normal structures. The non-local Coulomb corrections to conductances depend on the voltage difference $V_1-V_2$ not on their sum $V_1+V_2$, unlike in the superconducting case. All these features are directly related to the observation that only negative cross-correlations in shot noise occur in normal multi-terminal conductors \cite{BB}.

It is interesting to point out that S-like shaped non-local signal predicted here was indeed observed in experiments with NSN systems \cite{Beckmann2,newexp}. A good agreement between our theory and the results \cite{newexp} argues in favor of electron-electron interactions as a physical reason for the observed feature. Some of the features similar to those predicted here have also been observed in experiments \cite{Chandrasekhar}. It would be interesting to perform more experiments in systems under consideration both in superconducting and normal states
and compare the results with our theoretical predictions.
Extending experimental investigations to the normal case would hopefully allow for better characterization of the system parameters as well as for
clearer demonstration of qualitative differences between normal and superconducting structures outlined above.

Finally, we would like to make one more remark. In some cases non-linearities in both local and non-local differential conductances
caused by Coulomb interactions may be combined with the zero bias anomalies resulting from the proximity-enhanced electron interference in diffusive normal leads \cite{VZK,HN,Z,GKZ}.
In this paper we disregarded this effect for the sake of simplicity. In practice it implies that here we considered
the system with weakly disordered or sufficiently thick normal leads and sufficiently resistive barriers.
If needed, zero-bias anomaly effects \cite{VZK,HN,Z,GKZ} can be included into our analysis in a straightforward manner.

\end{document}